\documentclass[aps,prl,preprint,tightenlines,superscriptaddress,showpacs,byrevtex]{revtex4}
\usepackage{graphicx} 
\usepackage{dcolumn}  
\usepackage{epsfig}

\graphicspath{{ps}}

\begin{document}

\preprint{\vbox{ \hbox{   }
                 \hbox{BELLE-CONF-0643}
}}

\title
{\quad\\[0.5cm] Observation of a new $D_{sJ}$ meson in
$B^{+} \to \bar{D}^{0} D^{0} K^{+}$ decays}

\affiliation{Budker Institute of Nuclear Physics, Novosibirsk}
\affiliation{Chiba University, Chiba}
\affiliation{Chonnam National University, Kwangju}
\affiliation{University of Cincinnati, Cincinnati, Ohio 45221}
\affiliation{University of Frankfurt, Frankfurt}
\affiliation{The Graduate University for Advanced Studies, Hayama} 
\affiliation{Gyeongsang National University, Chinju}
\affiliation{University of Hawaii, Honolulu, Hawaii 96822}
\affiliation{High Energy Accelerator Research Organization (KEK), Tsukuba}
\affiliation{Hiroshima Institute of Technology, Hiroshima}
\affiliation{University of Illinois at Urbana-Champaign, Urbana, Illinois 61801}
\affiliation{Institute of High Energy Physics, Chinese Academy of Sciences, Beijing}
\affiliation{Institute of High Energy Physics, Vienna}
\affiliation{Institute of High Energy Physics, Protvino}
\affiliation{Institute for Theoretical and Experimental Physics, Moscow}
\affiliation{J. Stefan Institute, Ljubljana}
\affiliation{Kanagawa University, Yokohama}
\affiliation{Korea University, Seoul}
\affiliation{Kyoto University, Kyoto}
\affiliation{Kyungpook National University, Taegu}
\affiliation{Swiss Federal Institute of Technology of Lausanne, EPFL, Lausanne}
\affiliation{University of Ljubljana, Ljubljana}
\affiliation{University of Maribor, Maribor}
\affiliation{University of Melbourne, Victoria}
\affiliation{Nagoya University, Nagoya}
\affiliation{Nara Women's University, Nara}
\affiliation{National Central University, Chung-li}
\affiliation{National United University, Miao Li}
\affiliation{Department of Physics, National Taiwan University, Taipei}
\affiliation{H. Niewodniczanski Institute of Nuclear Physics, Krakow}
\affiliation{Nippon Dental University, Niigata}
\affiliation{Niigata University, Niigata}
\affiliation{University of Nova Gorica, Nova Gorica}
\affiliation{Osaka City University, Osaka}
\affiliation{Osaka University, Osaka}
\affiliation{Panjab University, Chandigarh}
\affiliation{Peking University, Beijing}
\affiliation{University of Pittsburgh, Pittsburgh, Pennsylvania 15260}
\affiliation{Princeton University, Princeton, New Jersey 08544}
\affiliation{RIKEN BNL Research Center, Upton, New York 11973}
\affiliation{Saga University, Saga}
\affiliation{University of Science and Technology of China, Hefei}
\affiliation{Seoul National University, Seoul}
\affiliation{Shinshu University, Nagano}
\affiliation{Sungkyunkwan University, Suwon}
\affiliation{University of Sydney, Sydney NSW}
\affiliation{Tata Institute of Fundamental Research, Bombay}
\affiliation{Toho University, Funabashi}
\affiliation{Tohoku Gakuin University, Tagajo}
\affiliation{Tohoku University, Sendai}
\affiliation{Department of Physics, University of Tokyo, Tokyo}
\affiliation{Tokyo Institute of Technology, Tokyo}
\affiliation{Tokyo Metropolitan University, Tokyo}
\affiliation{Tokyo University of Agriculture and Technology, Tokyo}
\affiliation{Toyama National College of Maritime Technology, Toyama}
\affiliation{University of Tsukuba, Tsukuba}
\affiliation{Virginia Polytechnic Institute and State University, Blacksburg, Virginia 24061}
\affiliation{Yonsei University, Seoul}
  \author{K.~Abe}\affiliation{High Energy Accelerator Research Organization (KEK), Tsukuba} 
  \author{K.~Abe}\affiliation{Tohoku Gakuin University, Tagajo} 
  \author{I.~Adachi}\affiliation{High Energy Accelerator Research Organization (KEK), Tsukuba} 
  \author{H.~Aihara}\affiliation{Department of Physics, University of Tokyo, Tokyo} 
  \author{D.~Anipko}\affiliation{Budker Institute of Nuclear Physics, Novosibirsk} 
  \author{K.~Aoki}\affiliation{Nagoya University, Nagoya} 
  \author{T.~Arakawa}\affiliation{Niigata University, Niigata} 
  \author{K.~Arinstein}\affiliation{Budker Institute of Nuclear Physics, Novosibirsk} 
  \author{Y.~Asano}\affiliation{University of Tsukuba, Tsukuba} 
  \author{T.~Aso}\affiliation{Toyama National College of Maritime Technology, Toyama} 
% ...
\collaboration{The Belle Collaboration}

\noaffiliation

\begin{abstract}

We report the observation of a new $D_{sJ}$ meson produced in 
$B^{+} \to \bar{D}^{0} D_{sJ} \to \bar{D}^{0} D^{0} K^{+}$.
This state has a mass of $M=2715 \pm 11 ^{+11}_{-14}~\rm{MeV}/c^{2}$, 
a width $\Gamma = 115 \pm 20 ^{+36}_{-32} ~\rm{MeV}/c^{2}$
and a spin-parity $1^{-}$.
The results are based on an analysis of $449$ million $B\bar{B}$ events 
collected at the $\Upsilon(4S)$ resonance
in the Belle detector at the KEKB asymmetric 
energy $e^{+} e^{-}$ collider.

\end{abstract}


\maketitle

\tighten

{\renewcommand{\thefootnote}{\fnsymbol{footnote}}}
\setcounter{footnote}{0}

At the level of quark diagrams, the decay $B \to \bar{D}D K$ 
proceeds dominantly via the CKM-favored 
$\bar{b} \to \bar{c} W^{+} \to \bar{c}c\bar{s}$ transition. 
The transition amplitudes can be 
categorized as either external $W$~-
or internal $W$~-emission (color-suppressed) diagrams.
The decay $B^{+}\to \bar{D}^{0}D^{0} K^{+}$~\cite{cc}
can proceed through both types of diagrams,
with the naive expectation that
the internal $W$  contribution to the branching fraction
is suppressed relative to that of the external $W$ 
by a factor of nine.  
$B$-meson decays
to three-body $\bar{D}DK$ final states are 
a promising area for searches for new
$c\bar{s}$ states as well as of some
$c\bar{c}$ states lying above $D^{0}\bar{D}^{0}$ 
threshold.   Since the externally emitted $W$ produces 
$1^{+},1^{-},0^{-}$ states, $c\bar{s}$ mesons
with these quantum numbers should be copiously produced.
The unexpected discoveries of the $D_{sJ}(2317)$
and  $D_{sJ}(2457)$ mesons show that our understanding of the
$c\bar{s}$ spectroscopy might be incomplete.
Experimental data on $c\bar{c}$ states with decay channels open
to $D^{(\star)}\bar{D}^{(\star)}$ are scarce.

The decays $B \to \bar{D}D K$ have been previously studied with a small 
data sample by 
ALEPH~\cite{aleph}. 
Recently a comprehensive study was performed by BaBar~\cite{babar}.
In this letter we report the 
first study of the
Dalitz plot of $B^{+}\to \bar{D}^{0} D^{0} K^{+}$ decay.
 
The study is performed using data 
collected  with the Belle detector at the KEKB asymmetric-energy
$e^+e^-$ (3.5 on 8~GeV) collider~\cite{KEKB},
operating at the $\Upsilon(4S)$ resonance ($\sqrt{s}=10.58$~GeV) 
with a peak luminosity that exceeds
$1.6\times 10^{34}~{\rm cm}^{-2}{\rm s}^{-1}$.
The data sample corresponds to the integrated luminosity
of $414~{\rm fb}^{-1}$ and contains 
$449$ million $B\bar{B}$ pairs. 

The Belle detector is a large-solid-angle magnetic
spectrometer that
consists of a silicon vertex detector,
a 50-layer central drift chamber (CDC), an array of
aerogel threshold \v{C}erenkov counters (ACC), 
a barrel-like arrangement of time-of-flight
scintillation counters (TOF), and an electromagnetic calorimeter
comprised of CsI(Tl) crystals (ECL) located inside 
a super-conducting solenoid coil that provides a 1.5~T
magnetic field.  An iron flux-return located outside of
the coil is instrumented to detect $K_L^0$ mesons and to identify
muons.  The detector
is described in detail elsewhere~\cite{Belle}.

We select charged tracks that originate from the interaction region
by requiring $\mid \! dr \! \mid < 0.4~\rm{cm}$ 
and $\mid \! dz \! \mid < 5~\rm{cm}$,
where $dr$ and $dz$ are the distances of closest approach to
the interaction point in the plane perpendicular to the beam 
and along the beam axis, respectively.
Charged particles are identified by using combined information
from the TOF, ACC and $dE/dx$ measurements in the CDC. 
Requirements on the particle identification variable are imposed
that identify a charge kaon with $90\%$ efficiency at $< 10\%$
$\pi \to K$ miss-identification probability and a charged pion with
$\approx 100\%$ efficiency at $< 10\%$ $K \to \pi$ miss-identification
probability. 
Any track that is positively identified as an electron is rejected.  

Candidate $K^{0}_{S} \to \pi^{+}\pi^{-}$ decays are identified
by a displaced secondary vertex, a two-pion momentum vector that is
consistent with a $K^{0}_{S}$ originating from the interaction
point and by the invariant mass selection 
$\mid \! M_{\pi^{+}\pi^{-}} - m_{K^{0}_{S}} \! \mid < 15~\rm{MeV/c^{2}}$.
Candidate $\pi^{0}$ mesons are identified as pairs 
of ECL-identified photons, each with a minimum energy of $50~\rm{MeV}$,
that have an invariant mass within $\pm 15~\rm{MeV/c^{2}}$
of the $\pi^{0}$ mass.

$D^{0}$ mesons are reconstructed in the
$K^{-}\pi^{+}$, $K^{-}\pi^{+}\pi^{+}\pi^{-}$, $K^{-}\pi^{+}\pi^{0}$,
$K^{0}_{S}\pi^{+}\pi^{-}$ and 
$K^{-}K^{+}$ 
decay modes.
We preselect $D$ candidates using a
signal window $\pm 30~\rm{MeV/c^{2}}$ 
around the nominal $D$ meson mass for decay modes, except for
$D^{0} \to K^{-}\pi^{+}\pi^{0}$ decays, where 
a larger $\pm 50~\rm{MeV/c^{2}}$ signal window
is used.
Mass- and vertex-constrained fits are applied to all
$D$ meson candidates to improve their momentum resolution.

We retain events that have a kaon candidate 
and at least two candidate $D^{0}$ mesons, with allowed flavor combination
and with the $D^{0}$ momenta in the $\Upsilon(4S)$ rest frame 
('cms')
below the kinematical limit for $B^{+} \to \bar{D}^{0} D^{0} K^{+}$.
To suppress the continuum background ($e^{+}e^{-} \to q\bar{q}, q=u,d,s,c$)
we require
the ratio of the second to the zeroth Fox-Wolfram moments~\cite{fox-wolfram}
to be less than $0.3$.
Momenta of the secondaries from a $B$ meson candidate decay
are refitted to a common vertex with an interaction point constraint
that takes into
account the $B$ meson decay length.

The $B$ meson candidates are identified by their cms energy difference,
$\Delta E = \Sigma_{i}E_{i} - E_{\rm beam}$,
and their beam constrained mass,
$M_{\rm bc} = \sqrt{E^{2}_{\rm beam} - (\Sigma_{i}\vec{p}_i)^2}$, 
where $E_{\rm beam}=\sqrt{s}/2$
is the beam energy in the cms and $\vec{p}_i$ and $E_i$ are
the three-momenta and energies of the $B$ candidate's decay products.
For the subsequent analysis we select $B$ candidates with
$M_{\rm bc}>5.2$~GeV$/c^2$ and $-0.4~{\rm GeV} < \Delta E <0.3$~GeV.
Exclusively reconstructed $B^{+} \to \bar{D}^{0} D^{0} K^{+}$ signal events
have $M_{\rm bc}$ distributions that peak at the nominal $B$-meson mass; 
the $\Delta E$ distributions peak near zero. 

We employ a discriminator (likelihood ratio) 
based on the $D^{0}$ meson signal significance
to select the unique $B$ candidate in the event, defined as:
${\cal LR}(M_{D}) = \frac{S(M_{D})}{S(M_{D})+B(M_{D})}$,
where $S$ and $B$ are the signal and the background probabilities
respectively.
This discriminator
is determined from the data, for each $D^{0}$ decay mode separately, 
using a sample enriched in $B \to \bar{D} D K$ decays.
For events with multiple
$B \to \bar{D} D K$ candidates,
the product
${\cal LR}_B = {\cal LR}(M_{\bar{D}}) \times {\cal LR}(M_{D})$
is calculated and the candidate with the largest ${\cal LR}(B)$
is accepted. 
The solution with larger kaon identification likelihood
${\cal L}(K)$ is chosen if multiple kaon candidates 
are found accompanying the accepted $\bar{D}^0 D^{0}$
combination.

The  ${\cal LR}_B$ discriminator is also used to suppress
combinatoric backgrounds to $B^{+} \to \bar{D}^{0} D^{0} K^{+}$
and to 
enhance the
signal purity.

The $\Delta E$ and $M_{\rm bc}$ distributions 
for the $B^{+}\to \bar{D}^{0}D^{0} K^{+}$ 
decay candidates, selected with  ${\cal LR}_B > 0.01$
requirement, are shown in Fig.~{\ref{dembc}.
\begin{figure}[!h]      
\vspace{-0.1cm}
\begin{center}      
\includegraphics[width=0.33\textwidth,height=0.31\textwidth]{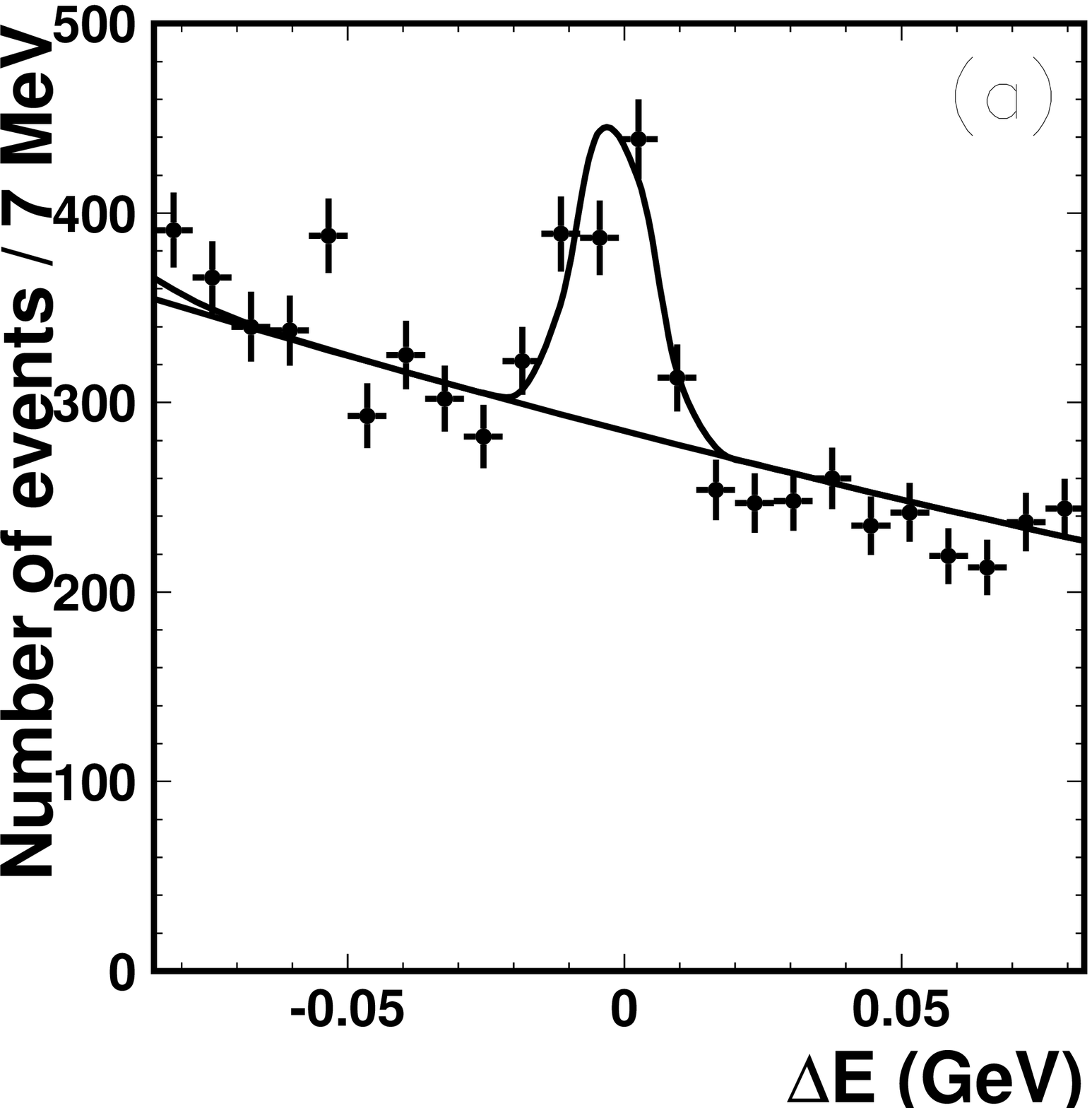}      
\includegraphics[width=0.33\textwidth,height=0.31\textwidth]{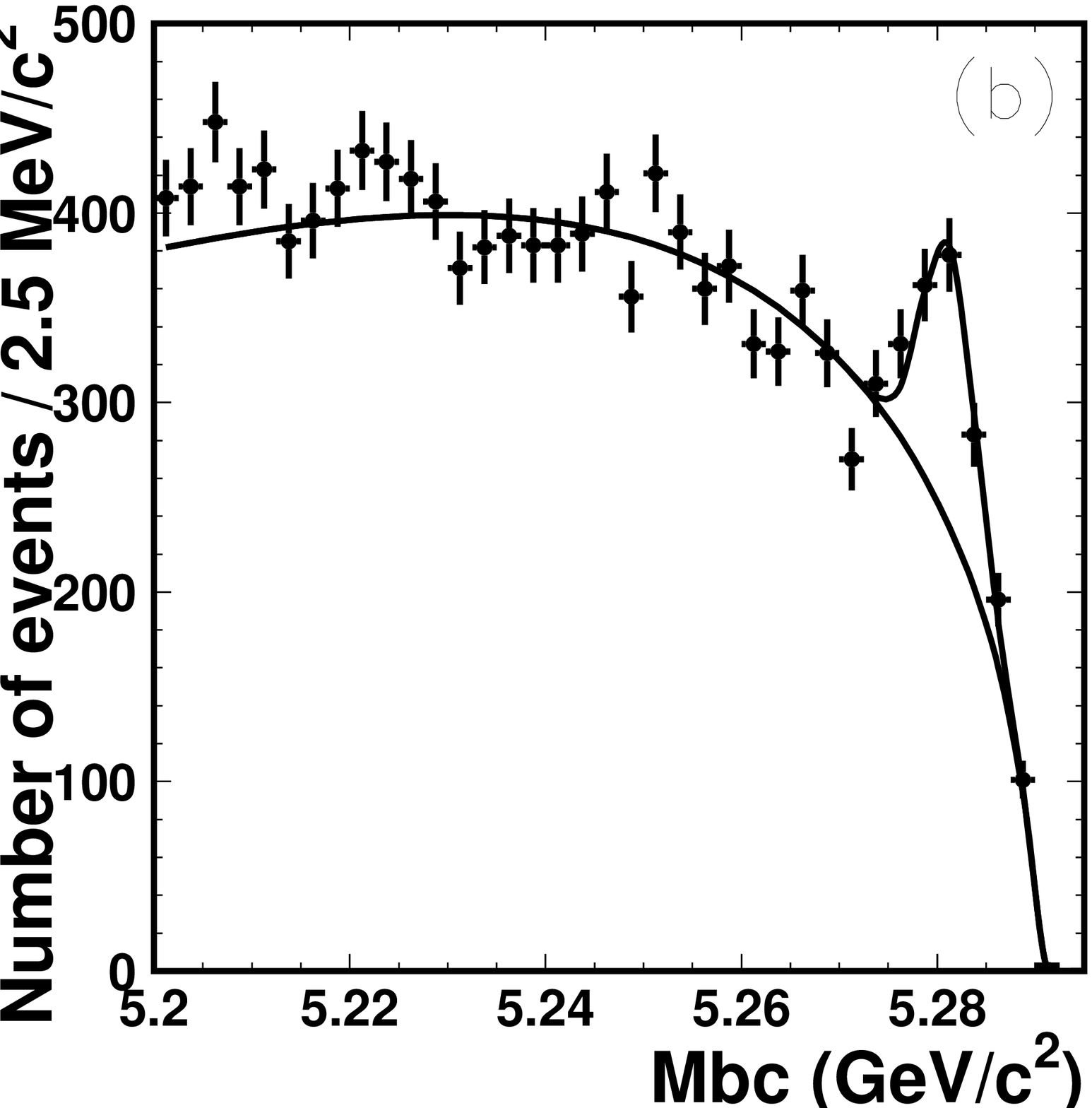}  
\vspace{-0.2cm}
\caption{$\Delta E$ (a) 
         and $M_{\rm bc}$ (b)
         distributions for
         $B^{+} \to \bar{D}^{0} D^{0} K^{+}$.}
\label{dembc}
\end{center}                                       
\vspace{-0.5cm}
\end{figure}      
The $\Delta E$ distribution is shown for events
with $\mid \!  M_{\rm bc} - m_{B} \!  \mid <  3\sigma_{M_{\rm bc}}$ 
($\sigma_{M_{\rm bc}} = 2.7~{\rm MeV}/c^{2}$), 
where $m_{B}$ is the nominal $B$ meson mass and
the $M_{\rm bc}$ distribution is shown for events with 
$\mid \! \Delta E \! \mid < 3\sigma_{\Delta E }$
($\sigma_{\Delta E } = 6.6~{\rm MeV}$).

From a study of the $M_{\rm bc}$ and $ \Delta E$ background distributions 
in large MC samples of generic $B\bar{B}$ and $q\bar{q}$ events, 
we find that the background level in the data is well explained by
the MC simulations and no peaking component is seen in  
either distribution.
Using events where one or both $D^{0}$ candidates 
are from the $D^0$-mass sidebands,
we verify with the data that there is no significnat
peaking background.

To extract the signal yield,
we perform extended unbinned maximum-likelihood fits simultaneously 
to $\Delta E$ and $M_{\rm bc}$.
The probability density functions (PDFs) for the $M_{\rm bc}$ and $\Delta E$ 
signals are Gaussians.
The background PDF for $M_{\rm bc}$ is represented by 
a phenomenological function~\cite{argus} with a phase-space-like
behaviour near the kinematic boundary; the
$\Delta E$ background is parameterized by a second-order polynomial.
The 
likelihood function is 
maximized
with free parameters for the signal yield, the Gaussian means
and widths, and four 
parameters that describe shapes of the background
distributions.

From the fit, we obtain a signal yield of $N_{\rm sig}= 399\pm 40$ 
events with a signal-to-background ratio of $S/B \simeq 0.3$.
The results of the fit are superimposed on
the $\Delta E$ and $M_{\rm bc}$ projections shown 
in Fig.~\ref{dembc}.
We determine the branching fraction from the relation
\begin{equation}
{\cal B}(B^{+}\to \bar{D}^{0}D^{0}K^{+})=
\frac{N_{sig}}{N_{B^{+}B^{-}}\sum_{ij}\epsilon_{ij}{\cal B}(\bar{D}\to i){\cal B}(D\to j)},
\label{bf}
\end{equation}
where $\epsilon_{ij}$ are efficiences for the $i$ and $j$ $D$ subchannels and
for $N_{B^{+}B^{-}}$; we assume 
$N_{B^{+}B^{-}} = N_{B^{0}\bar{B}^{0}}$.
The efficiences are determined by MC using a phase-space decay model.
The sum in the denominator of Eq.(\ref{bf}) amounts to $6.8\times 10^{-4}$.

We obtain
${\cal B}(B^{+}\to \bar{D}^{0}D^{0}K^{+})=(13.1\pm 1.3^{+1.7}_{-2.7})\times 10^{-4}$,
where the first error is statistical and the second is
systematic.  The latter includes contributions due to 
uncertainties in the efficiency determination
(tracking and particle identification efficiency, data-MC differences in 
$\Delta E, M_{\rm bc}$ signal shapes), 
the ${\cal LR}_B$ selection, the background parameterization, the
MC model used in the efficiency calculation, 
the intermediate $D \to {\rm i}$ branching fractions and $N_{B^{+}B^{-}}$. 

\begin{figure*}[!ht]      
\vspace{-0.1cm}
\begin{center}      
\includegraphics[width=0.33\textwidth,height=0.31\textwidth]{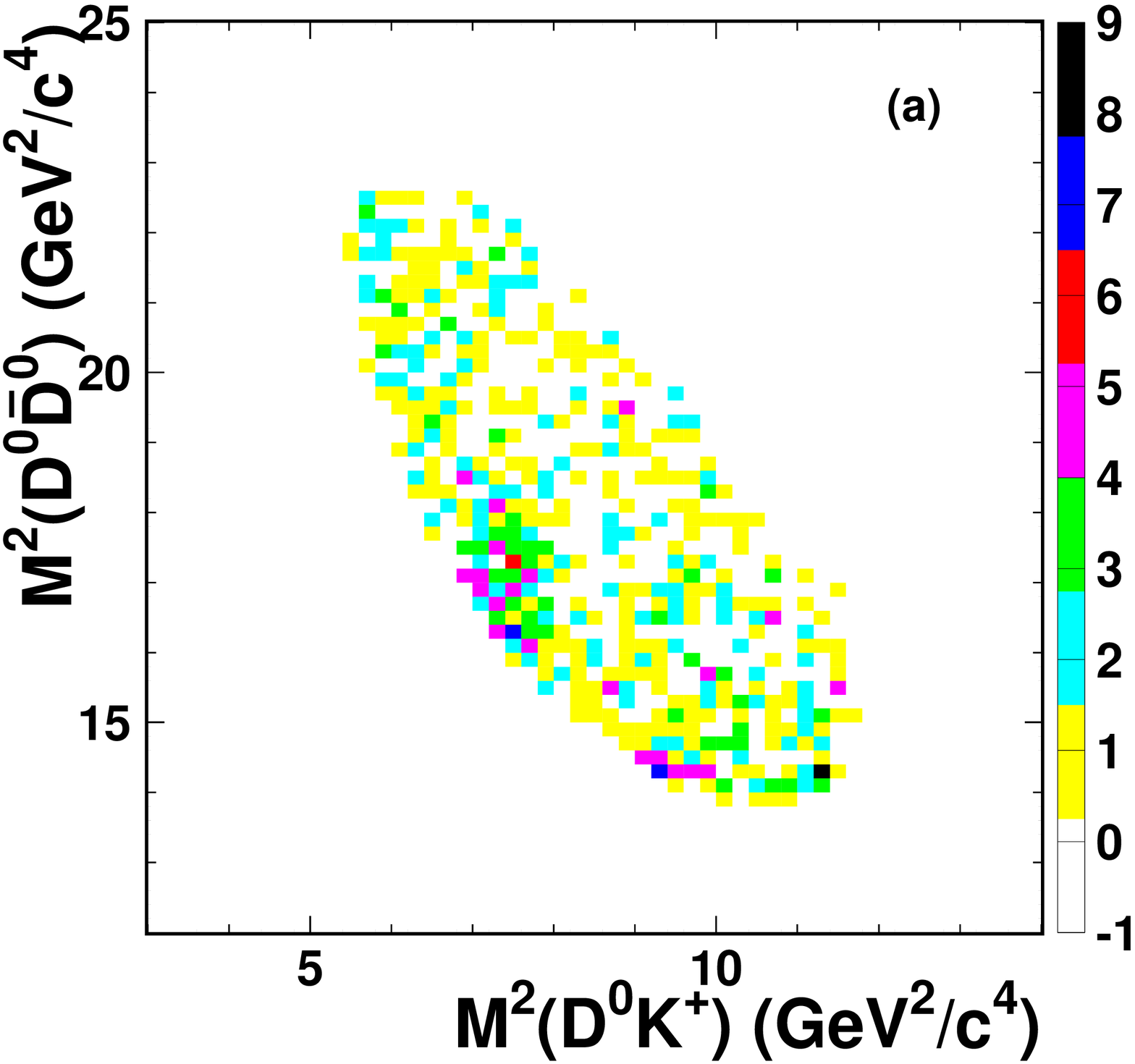}
\includegraphics[width=0.33\textwidth,height=0.31\textwidth]{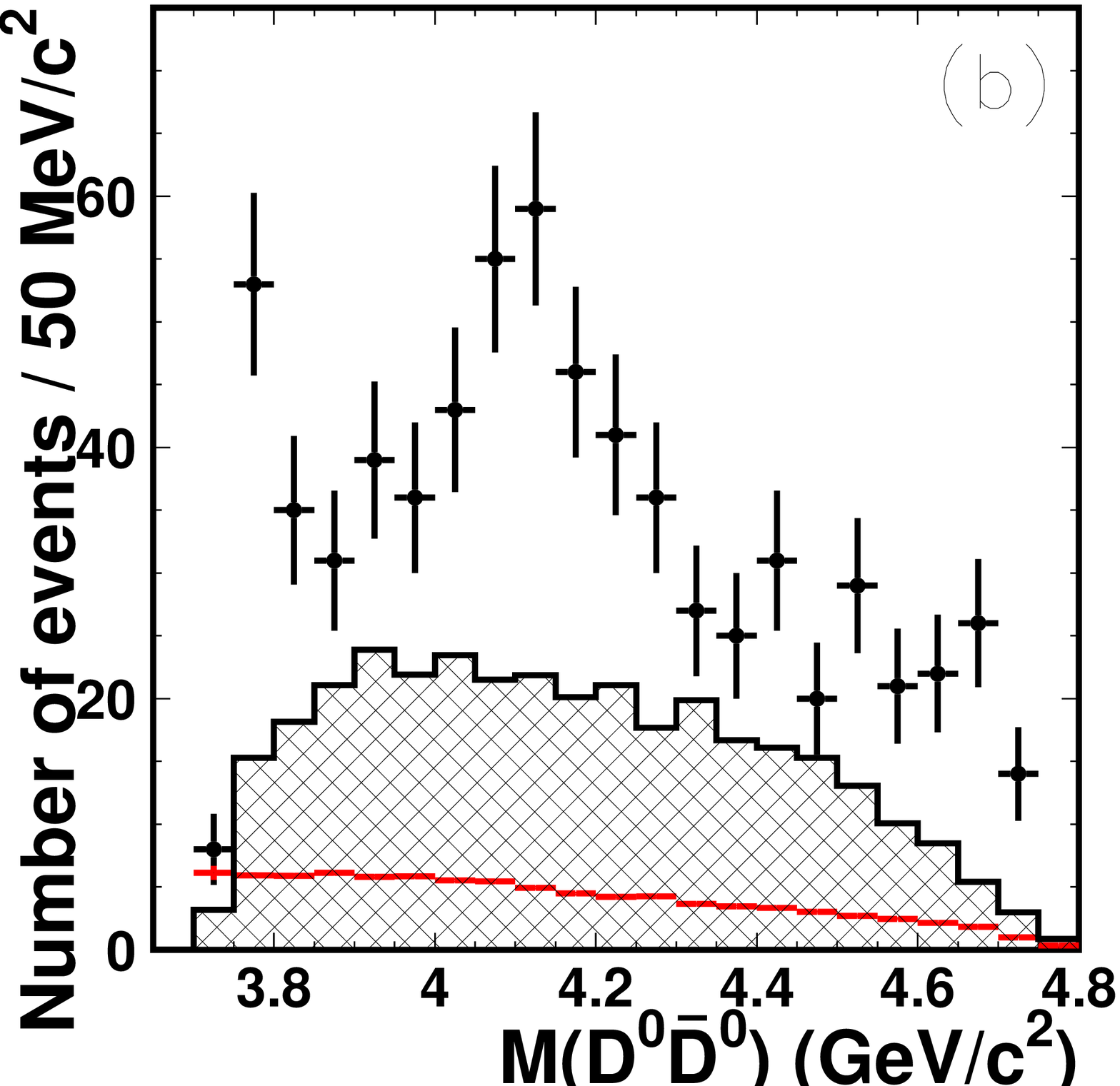}      
\includegraphics[width=0.33\textwidth,height=0.31\textwidth]{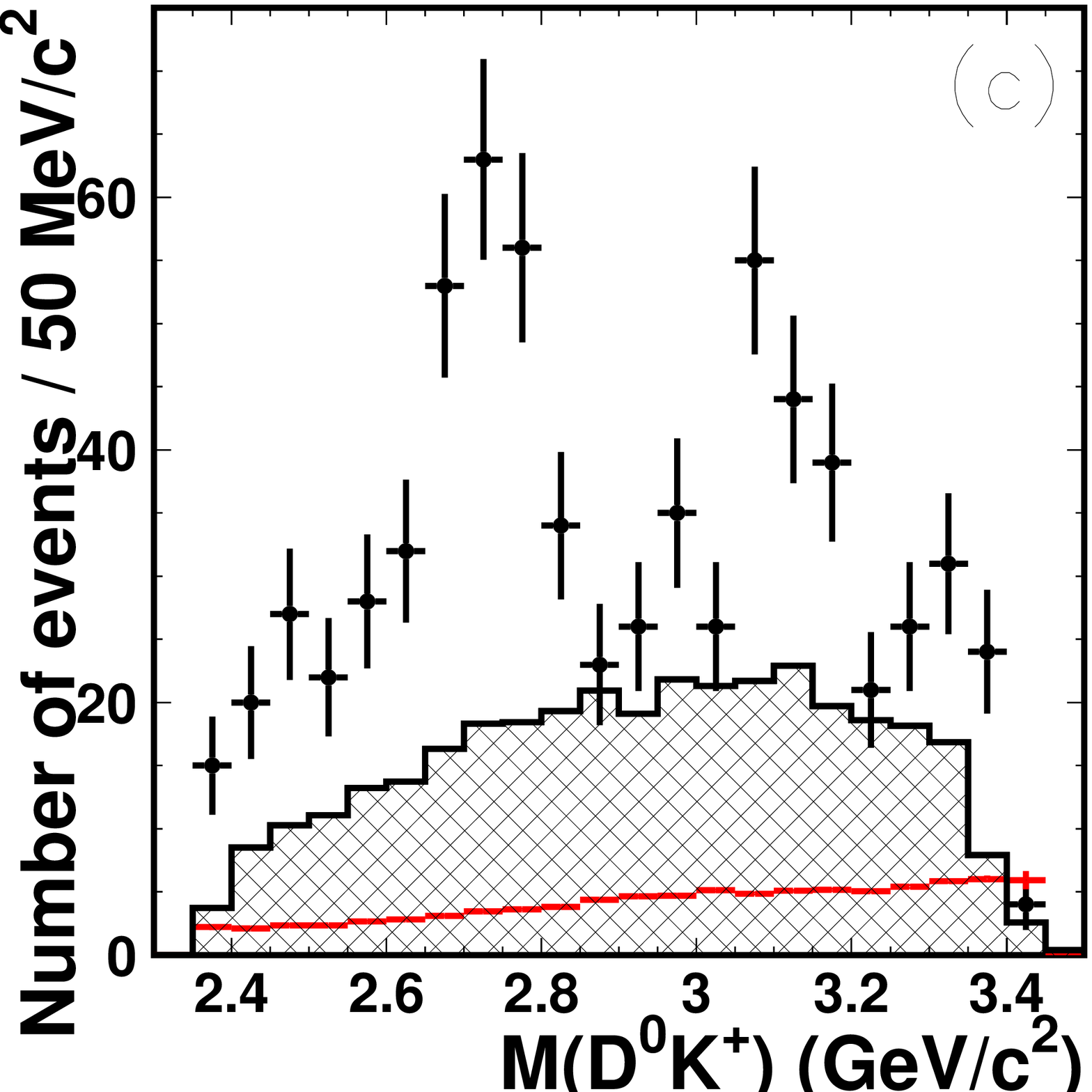}      
\includegraphics[width=0.33\textwidth,height=0.31\textwidth]{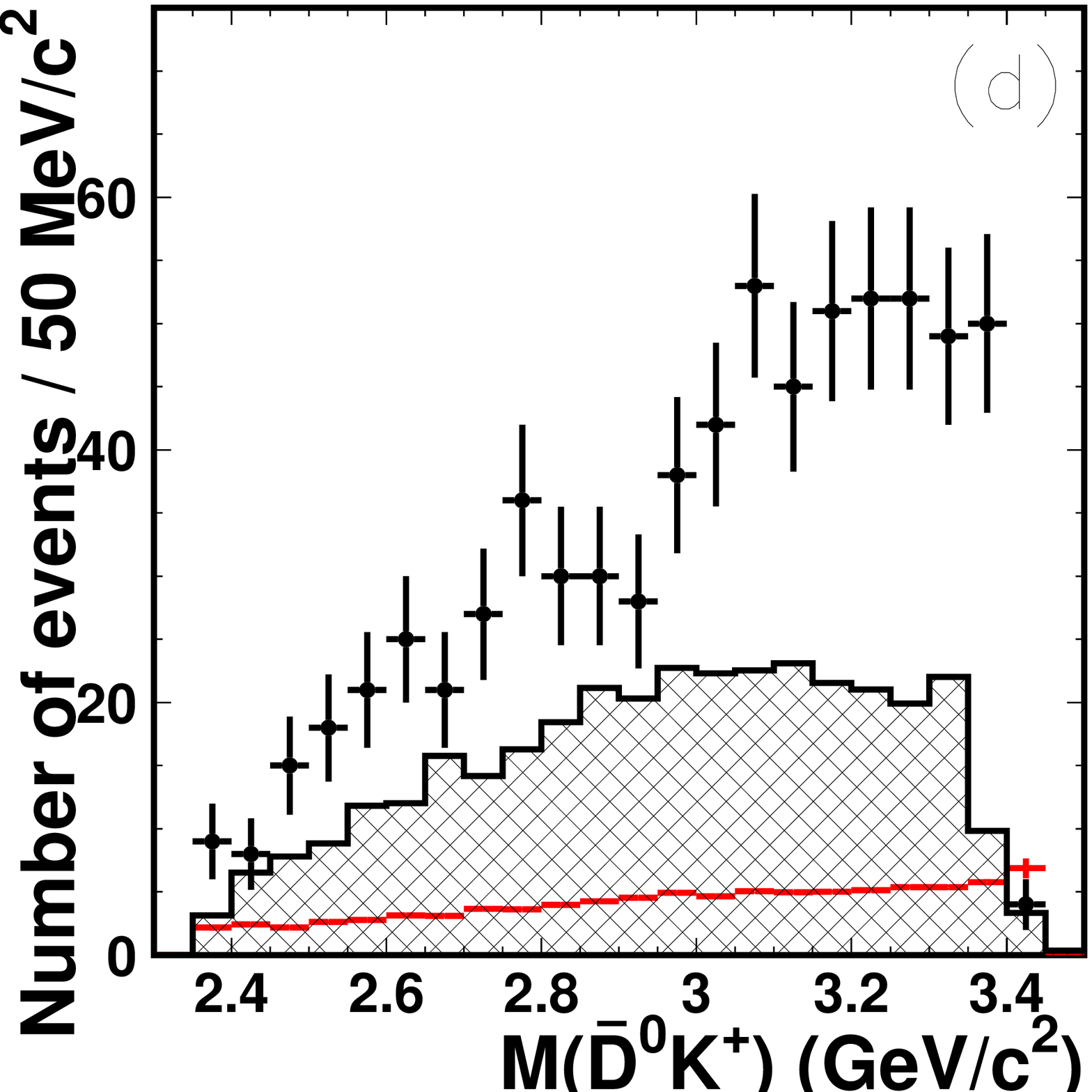}      
\vspace{-0.2cm}
\caption{ Dalitz plot (a) and projections for $B^{+}\to \bar{D}^{0} D^{0} K^{+}$
          in $1.5\sigma ~\Delta E \!- \! M_{\rm bc}$ signal region:
           $M(D^{0} \bar{D}^{0})$(b),
           $M(D^{0}K^{+})$(c), 
           $M(\bar{D}^{0}K^{+})$(d). 
Hatched histograms represent background,  
red curves show the efficiency variation (vertical scale in $\%$).}
\label{dalitz_d0d0bkc}
\end{center}                                       
\vspace{-0.6cm}
\end{figure*}   

The Dalitz plot 
$M^{2}(D^{0}\bar{D}^{0})$~vs~$M^{2}(D^{0}K^{+})$
for events from a signal region defined by the ellipse
$(\Delta E / 1.5\sigma_{\Delta E})^{2}+
((M_{\rm bc}-m_{B})/1.5\sigma_{M_{\rm bc}})^{2} < 1$
is shown in Fig.~\ref{dalitz_d0d0bkc}(a).
The three, two-body invariant mass distributions are 
shown in Figs.~\ref{dalitz_d0d0bkc}(b)-(d).
The hatched histograms 
represent the background distributions obtained for events
from an elliptical strip surrounding the $\Delta E$, $M_{\rm bc}$
signal region that extends from $6\sigma_{\Delta E},\sigma_{M_{\rm bc}}$
to $10\sigma_{\Delta E},\sigma_{M_{\rm bc}}$.
The background distributions are normalized to the number
of background events under the signal 
peaks $(\pm 1.5\sigma)$ as determined from the combined 
$\Delta E$ and $M_{\rm bc}$ fit.
The data shown are not efficiency corrected.
The variation of efficiency
as a function of invariant mass is shown in 
Figs.~\ref{dalitz_d0d0bkc}(b)-(d) as a continuous curve. 

A pronounced feature of the Dalitz plot 
is the accumulation of events in the region
$16~\rm{GeV}^{2}/c^{4} < M^{2}(D^{0} \bar{D}^{0}) < 18~\rm{GeV}^{2}/c^{4}$ 
and $7~\rm{GeV}^{2}/c^{4} < M^{2}(D^{0}K^{+}) < 8~\rm{GeV}^{2}/c^{4}$,
possibly the overlap of 
a horizontal band that could be due to the $\psi(4160)$   
and a vertical band that
cannot be attributed to any known $c \bar{s}$ state.
A horizontal band at $M^{2}(D^{0}\bar{D}^{0}) \simeq 14.2~\rm{GeV}^{2}/c^{4}$ 
corresponds to the $\psi(3770)$ production.  

We employ simultaneous fits to the $\Delta E$ and $M_{\rm bc}$ distributions 
for events from each $50~\rm{MeV}/c^{2}$ mass bin of the Dalitz plot projection  
to obtain background-subtracted invariant mass distributions.
In these fits the widths and positions of the Gaussians describing
the signal are fixed at
the values obtained for the total signal sample, while
the signal yield 
and the background
PDF's paremeters are free parameters.
The obtained 
signal yields are shown in Fig.~\ref{xx_d0d0bkc_plus_reflections}
as points with error bars. 

The $\psi(3770)$ signal is studied with  
finer, $20~\rm{MeV}/c^{2}$ mass bins (Fig.~\ref{d0d0k_mass_bins}(a)).
The peak is fitted in the region 
$M(D^{0} \bar{D}^{0}) < 4~\rm{GeV}/c^{2}$ 
with a Breit-Wigner (BW) plus a threshold
function to describe a
nonresonant component. The $\psi(3770)$ signal yield is $68 \pm 15$
events with a peak mass of $3777\pm 3~\rm{MeV}/c^{2}$, 
and a width of $27 \pm 9~\rm{MeV}/c^{2}$,
in agreement with the PDG averages.

\begin{figure}[!hb]      
\vspace{-0.3cm}
\begin{center}      
\includegraphics[width=0.328\textwidth,height=0.31\textwidth]{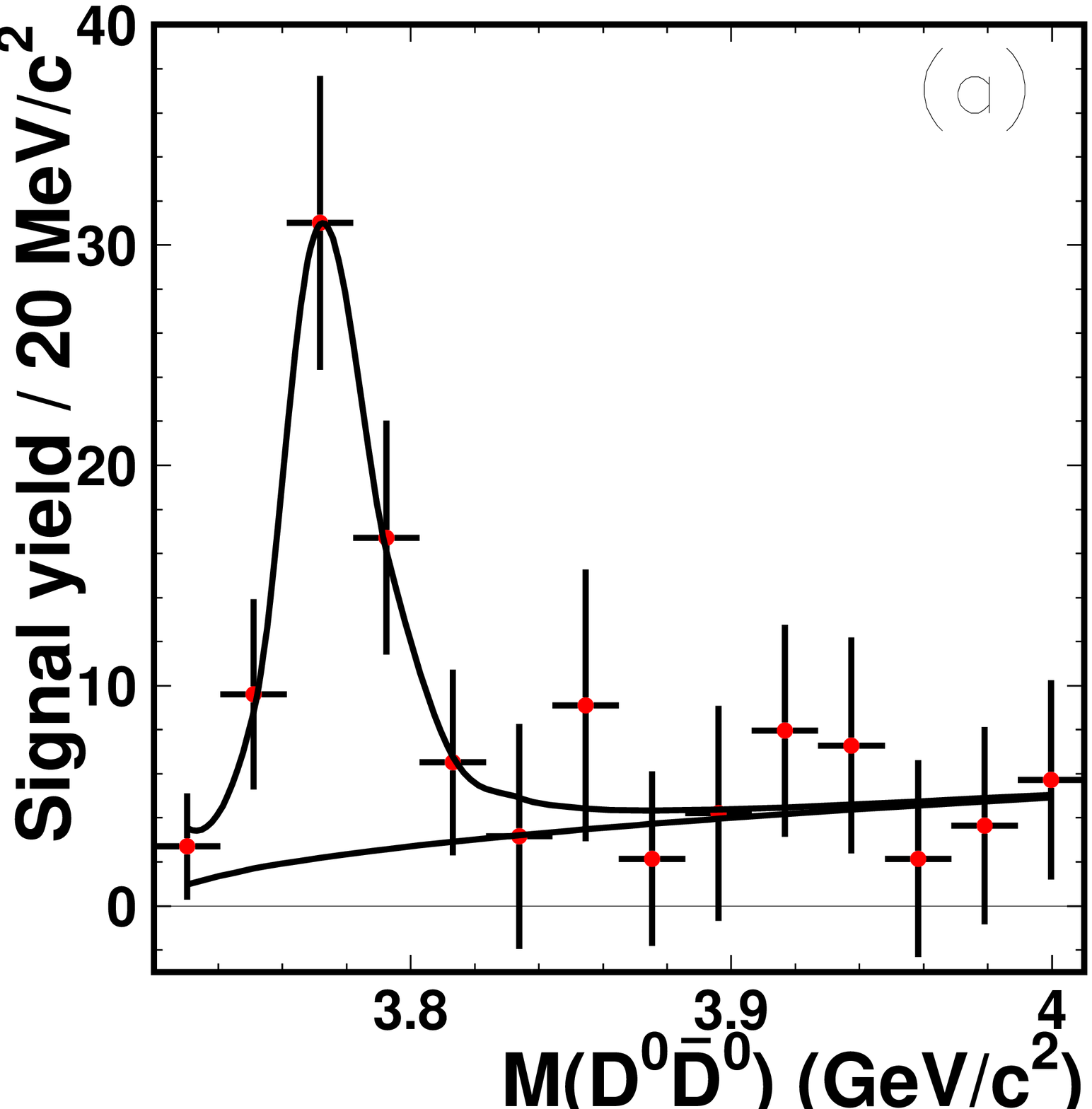}      
\includegraphics[width=0.328\textwidth,height=0.31\textwidth]{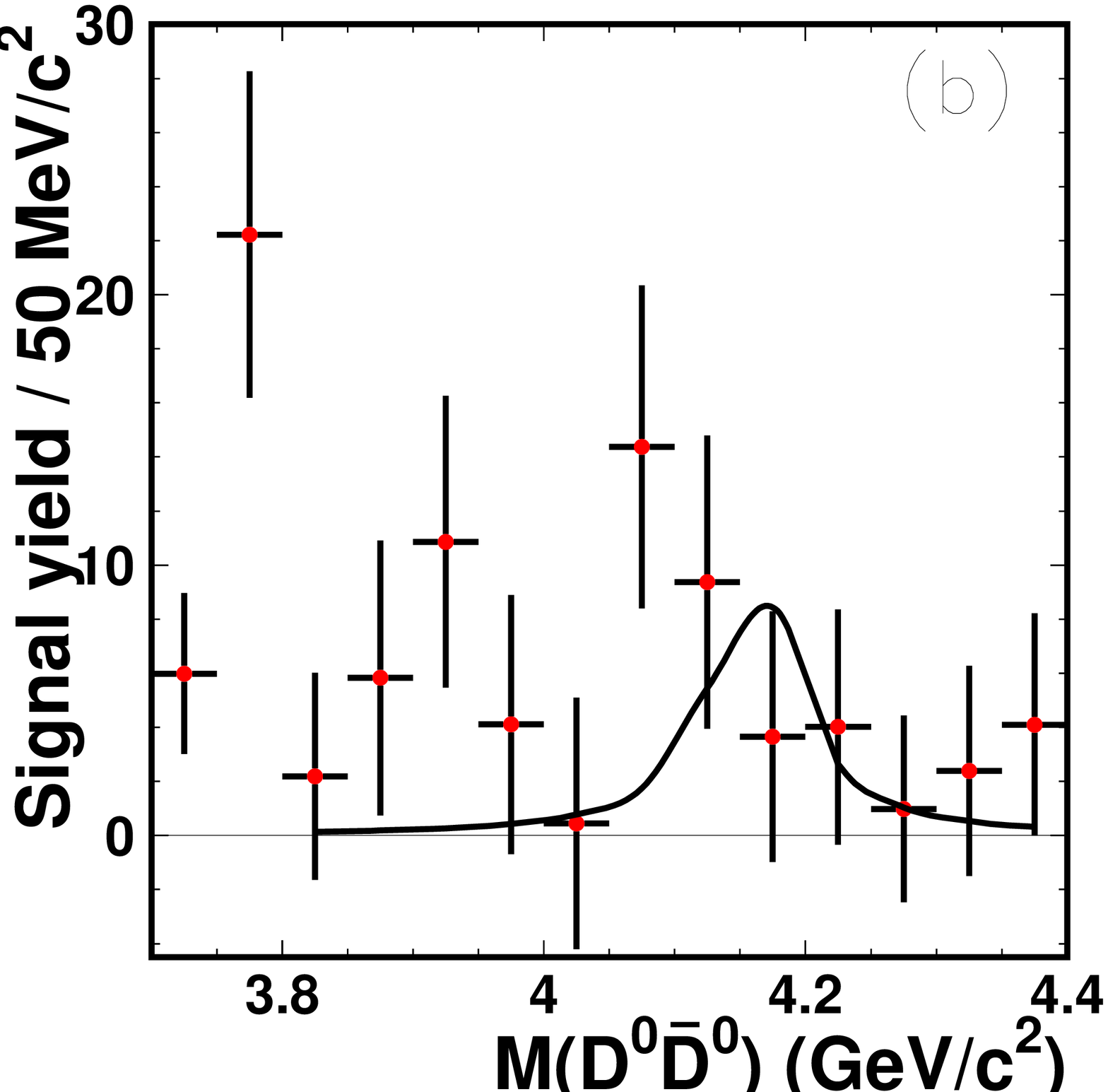}      
\includegraphics[width=0.328\textwidth,height=0.31\textwidth]{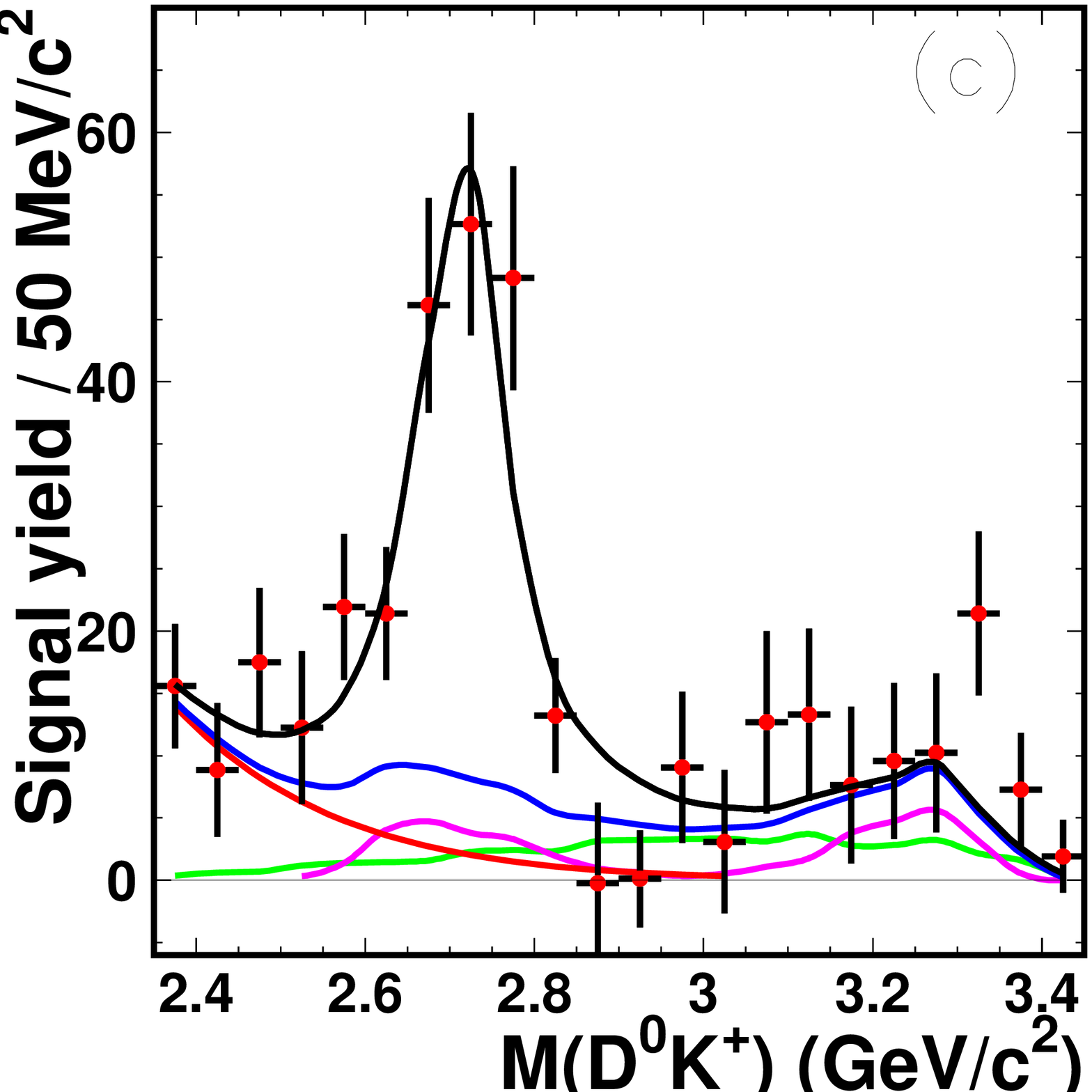}    
\vspace{-0.2cm}
\caption{Background-subtracted mass distributions:
         (a) $\psi(3770)$ signal, 
         (b) $M(D^{0} \bar{D}^{0})$ for 
             $cos\theta_{hel}>0$
         (c) $M({D}^{0}K^{+})$ for $M(D^{0}\bar{D}^{0}) > 3.85~\rm{GeV}/c^{2}$
Solid curves denote fit results described in the text. 
The lower curve in (a) shows phase-space component, 
coloured curves in (c): purple - the $\psi(4160)$ reflection, green: 
phase-space,
red: threshold component and navy-blue: the 
sum of the three components. 
         }
\label{d0d0k_mass_bins}
\end{center}      
\vspace{-0.6cm}
\end{figure}
The background-subtracted 
$M(D^{0}\bar{D}^{0})$ spectrum 
(Fig.~\ref{d0d0k_mass_bins}(b)),
for events 
satisfying $\cos\theta_{hel} > 0$, 
where $\theta_{hel}$ is the helicity angle
between the $D^{0}$ momentum vector and the boost direction to
the $D^{0} \bar{D}^{0}$ rest frame,
is used to estimate the possible $\psi(4160)$    
contribution to the enhancement 
at $M(D^{0}K^{+}) \simeq  2.7~\rm{GeV}/c^2$.
The distribution for $M(D^{0}\bar{D}^{0}) > 3.8~\rm{GeV}/c^{2}$
is fitted with a BW with mass and width 
fixed at the nominal $\psi(4160)$ values 
($M=4160$, $\Gamma = 80~\rm{MeV}/c^{2}$~\cite{pdg04}),  
yielding $24 \pm 11$
signal events.    We use these $\psi(4160)$ parameters 
to estimate the number of $\psi(4160)$ events in the backward 
helicity-angle hemisphere, in the region
$M({D}^{0}K^{+}) < 2.9~\rm{GeV}/c^{2}$.

Figure~\ref{d0d0k_mass_bins}(c) shows the
background-subtracted $M({D}^{0}K^{+})$
distribution for events with $M(D^{0}\bar{D}^{0}) > 3.85~\rm{GeV}/c^{2}$.
This requirement removes the $\psi(3770)$ reflection at 
high $M({D}^{0}K^{+})$.
The predicted  $\psi(4160)$ reflection is indicated in the figure  
by the purple curve. 
The $\psi(4160)$ reflection agrees well with the data in the high mass
$M({D}^{0}K^{+})$ region but does not explain the large
peak at $M({D}^{0}K^{+})\simeq 2.7~\rm{GeV}/c^{2}$.
We parameterize the observed excess of events with a BW
and fit the $M({D}^{0}K^{+})$ spectrum (Fig.~\ref{d0d0k_mass_bins}(c))
with the ansatz of a new resonance,
the $\psi(4160)$ reflection shape and a 
phase-space component as determined by MC simulations.
The free parameters in the fit are the resonance yield, mass and width, 
and the phase-space component normalization.
The fit has an acceptable overall $\chi ^2$ but is unable to reproduce
the events near the low-mass threshold seen in Fig.~\ref{d0d0k_mass_bins}(c).
We used several phenomenological 
parameterizations
(polynomials, another BW, an exponential) of the threshold enhancement
into the fit to determine its influence on the BW parameters 
of the $2.7~\rm{GeV}/c^{2}$ peak.
The exponential form $a\times exp{[-\alpha M^2(D^{0}K^{+})]}$ 
gives a good description of the mass spectrum, while adding only two
free parameters.
From this fit
we obtain for this new resonance, which we further denote 
as the  $D_{sJ}^{+}(2700)$,
a signal yield of $182 \pm 30$ events, mass of 
$M = 2715 \pm 11~\rm{MeV}/c^{2}$ and width of 
$\Gamma = 115 \pm 20~\rm{MeV}/c^{2}$. 
The threshold and the phase-space components from the
fits are $58 \pm 38$ and $47\pm 26$ events, respectively.
The fit results are depicted in 
Figs.~\ref{xx_d0d0bkc_plus_reflections}(a)-(c)
as histograms overlaid on the measured mass spectra.

The resonance parameters and
product branching fractions 
are summarized in Table~\ref{bf_table}.
The $\psi(4160)$ yield is not statistically significant, 
and therefore a $90\%$ C.L. upper limit is also quoted.
The ${\cal B}$'s of the threshold and the phase-space
components are $(1.9  \pm 1.2 ^{+1.0} _{-1.1})\times 10^{-4}$
and $(1.5 \pm 0.8^{+0.2}_{-1.6})\times 10^{-4}$ 
(the first errors are statistical, the second are systematic)
which correspond to
the $90\%$ C.L. upper limits of
$4.9 \times 10^{-4}$ and $3.0 \times 10^{-4}$, respectively.
The systematic errors on the product branching fractions 
and the resonance parameters
include contributions from 
the efficiency variation over the Dalitz plot,  
uncertaintities in the yields of the $\psi(4160)$ reflection 
(including recent measurements of the $\psi(4160)$ parameters~\cite{bes_cb}), 
the threshold parameterization,  
sensitivities of parameters to the fit range and 
parameterization, uncertainties in the ${\cal LR}_B$ selection, 
as well as due to neglected interference effects. 
The systematics due to the latter  are
determined from MC simulations of Dalitz plot densities
with and without interference of coherent amplitudes.  
Here an isobar formalism was used, with each contributing resonance
parameterized by the BW form. 
The resonance parameters from Table~\ref{bf_table}
and the threshold enhancement parameters are used to 
determine the amplitudes.
The effects of interference of the $\psi(3770)$ with other states are found
small
and are neglected in the simulations.
Events generated with maximal constructive and destructive interferences
between the amplitudes, were passed through a complete detector
simulation and analysed ignoring interference effects.
The differences between the obtained resonance parameters
and the input values are taken as their systematic errors.

\begin{table}[]
\vspace{-0.1cm}
\caption[] { Branching fractions of quasi-two-body components}
\begin{tabular}{l|c |c |c } 
$    R         $          & $D_{sJ}^{+}(2700)$ 
                         & $\psi(3770)$
                         & $\psi(4160)$
                                                 \\ \hline \hline
${\rm N_{sig}}$            & $ 182 \pm 30$    
                         & $ 68 \pm 15$
                         & $ 43 \pm 20 $       
                                                 \\ \hline
${\rm M (MeV/c^{2})}$      & $ 2715 \pm 11^{+11}_{-14}$
                         & $ 3777 \pm 3  \pm 4$ 
                         & $ 4160 (\rm{fixed})$           
                                                 \\ \hline
${\rm \Gamma (MeV/c^{2})}$& $ 115 \pm 20^{+36}_{-32}$
                         & $ 27 \pm 9 \pm 5$ 
                         & $ 80 (\rm{fixed})$
                                                 \\ \hline
${\cal B}[B^{+} \! \to R K^{+} (\bar{D}^{0} R) ] \times$     
                         & $ 7.2$  
                         & $ 1.5$  
                         & $ 1.1$  \\
${\cal B}[R \! \to \bar{D}^{0}D^{0} (D^{0}K^{+})]$    
                         &$\pm 1.2^{+1.0}_{-2.9} $
                         &$\pm 0.3 ^{+0.2}_{-0.3}$  
                         &$\pm 0.5^{+0.5}_{-0.2} $ \\
$[10^{-4}] \rm{(or} ~$90\%$ ~\rm{C.L.)} $  & &  & $(<2.4)$\\ \hline \hline
\end{tabular}
\label{bf_table}
\vspace{-0.2cm}
\end{table}
We study background-subtracted 
$\psi(3770)$  and $D_{sJ}^{+}(2700)$ helicity angle distributions
by selecting the respective invariant mass in the resonance
region and obtaining signal yields in bins of $\cos\theta_{hel}$
from simultaneous fits to $\Delta E$ and $M_{\rm bc}$. Spin-parity
hypotheses for the resonances are tested by performing
binned $\chi^{2}$ fits to the obtained angular distributions
corrected with efficiency weights.
The $J=1$ hypothesis describes the $\psi(3770)$ 
distribution well  
($\chi^{2}/ndf=3.6/5$),
$J=1$ is favoured for the $D_{sJ}^{+}(2700)$ ($\chi^{2}/ndf=7/5$);
the $J=0$ ($185/5$) and $J=2$ ($250/5$) assignments can be rejected.
\begin{figure}[!h]      
\vspace{-0.1cm}
\begin{center}      
\includegraphics[width=0.328\textwidth,height=0.31\textwidth]{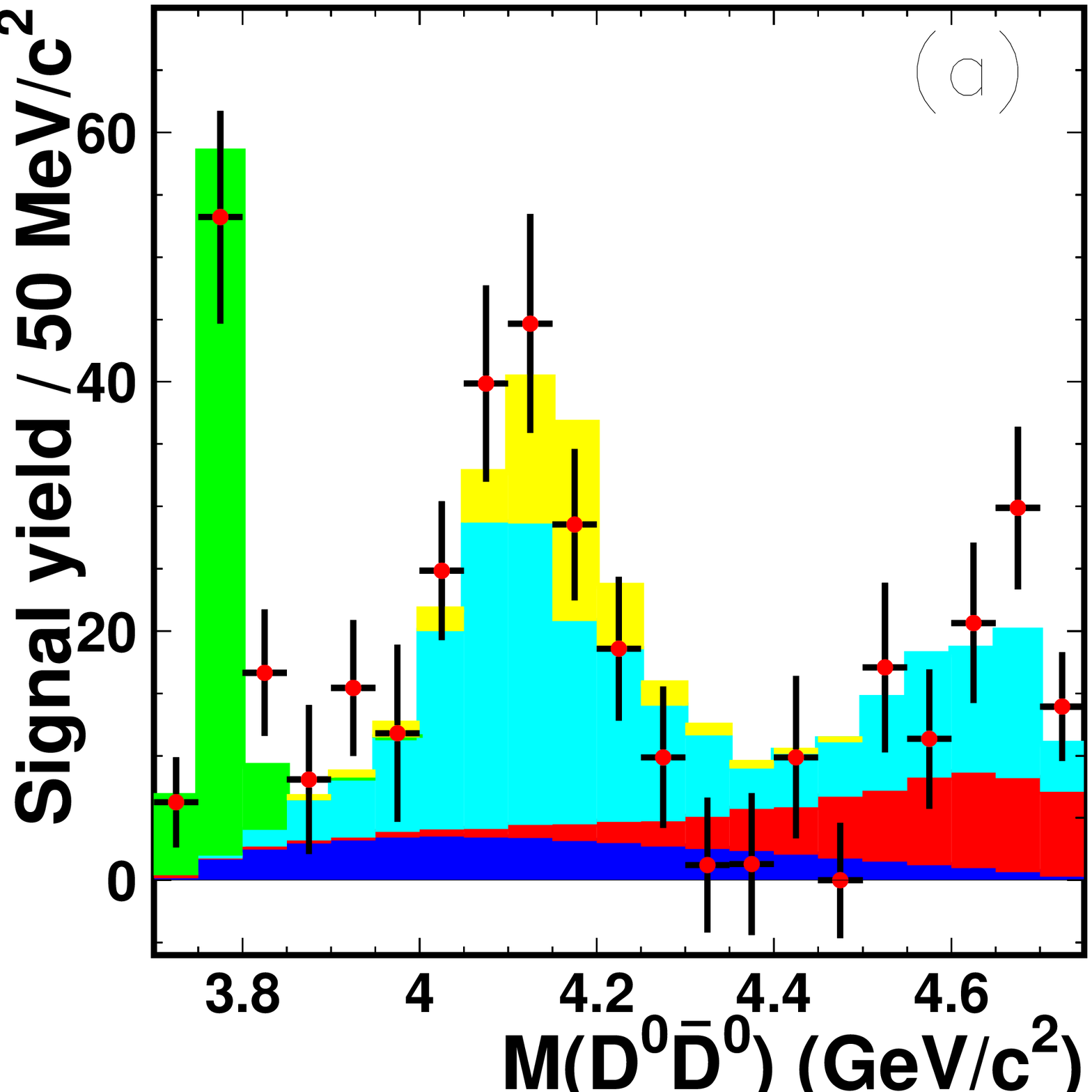}      
\includegraphics[width=0.328\textwidth,height=0.31\textwidth]{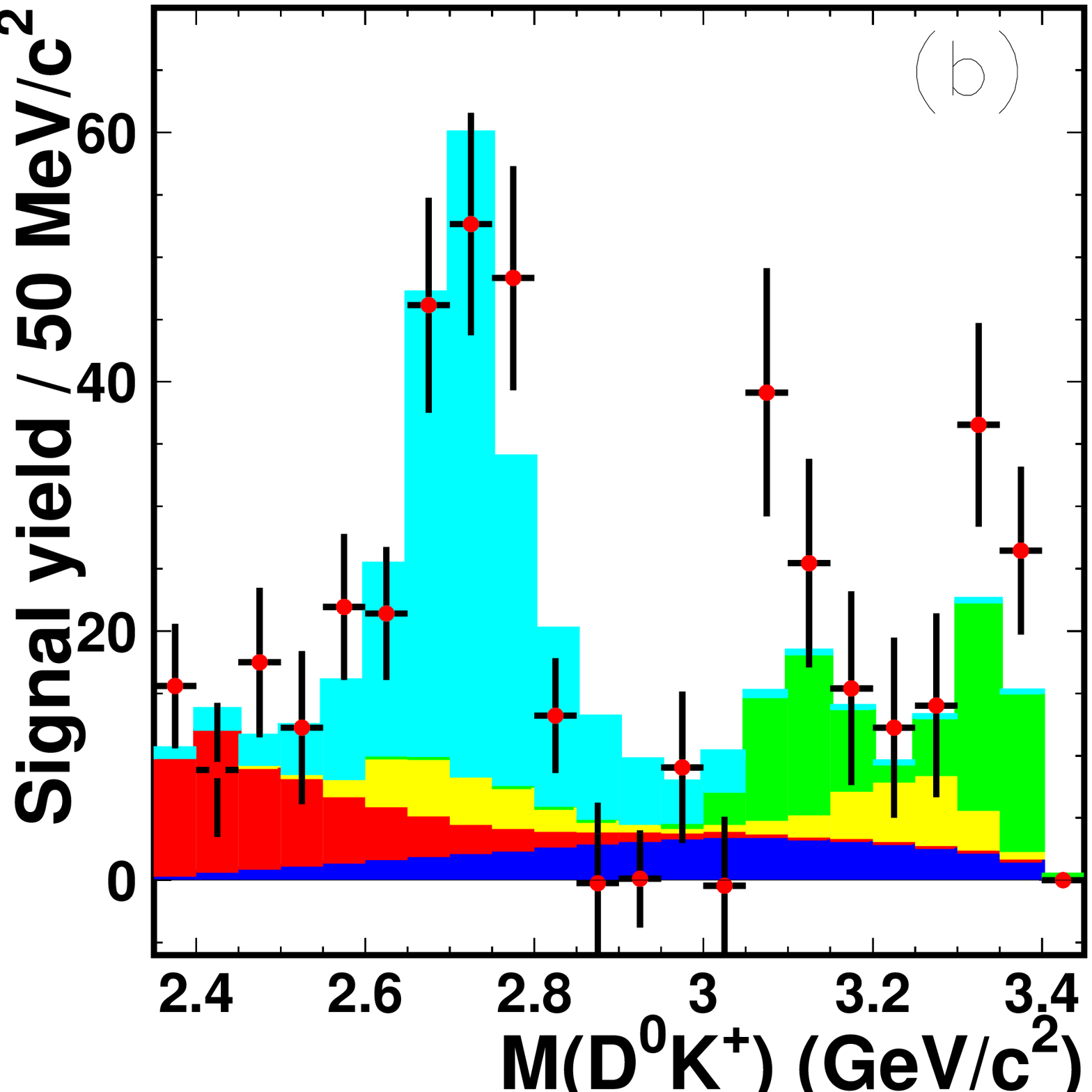}      
\includegraphics[width=0.328\textwidth,height=0.31\textwidth]{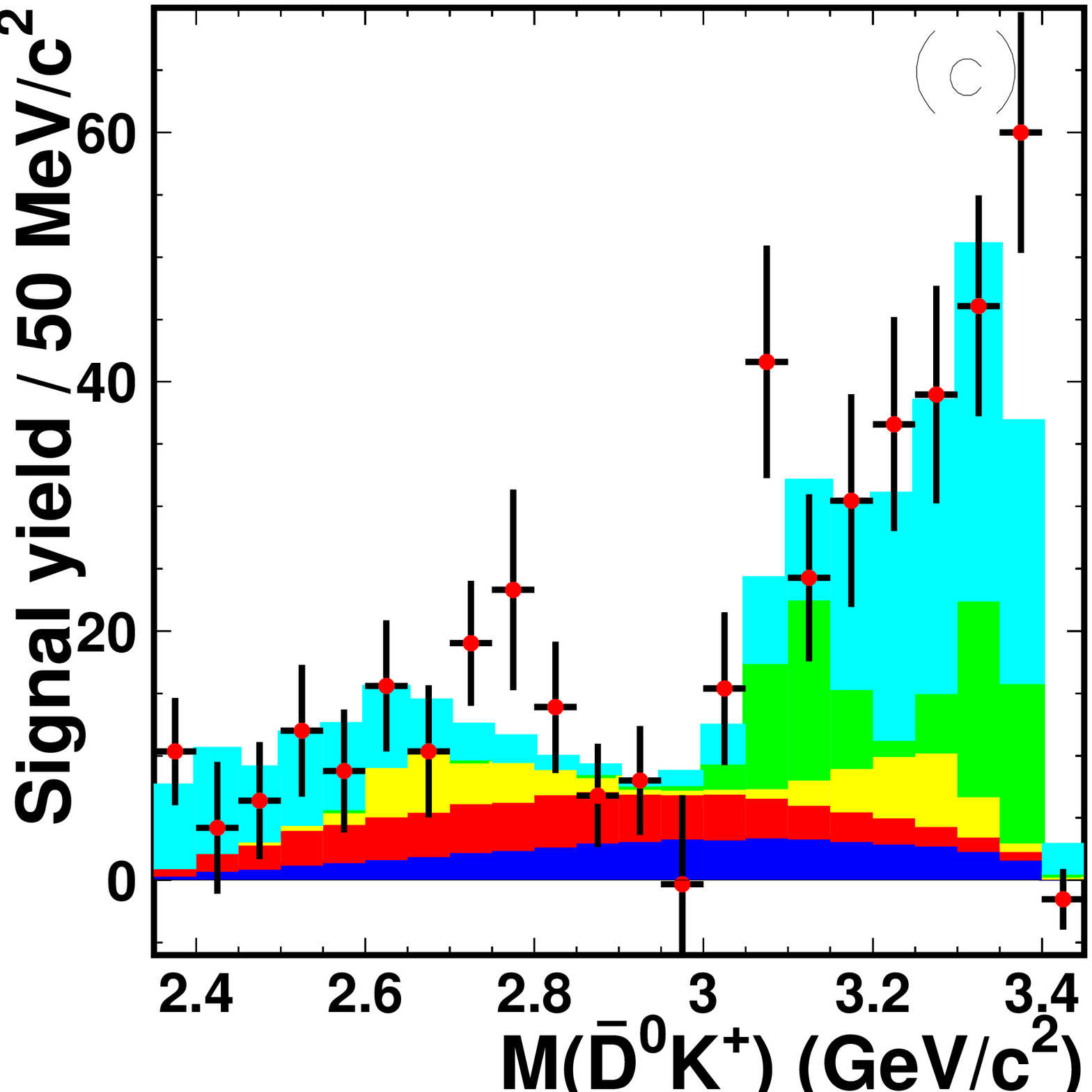}     
\vspace{-0.2cm}
\caption{$B^{+} \to \bar{D}^{0} D^{0} K^{+}$ signal yield vs
         $M(D^{0}\bar{D}^{0})$~(a), $M({D}^{0}K^{+})$~(b),
         and $M(\bar{D}^{0}K^{+})$~(c).
         Histograms denote the contributions from:
         $D_{sJ}^{+}(2700)$ (blue),
         $\psi(3770)$ (green), $\psi(4160)$ (yellow), 
         threshold (red) and  
         phase-space components (navy blue).
         Histograms are superimposed additively. }
\label{xx_d0d0bkc_plus_reflections}
\end{center}      
\vspace{-0.6cm}
\end{figure}      

The $M({D}^{0}K^{+})$ distribution of 
Fig.~\ref{dalitz_d0d0bkc}(c) in $10~\rm{MeV}/c^{2}$
is used to search for the $D_{sJ}(2573)$ contribution.
We include this state in the fit 
using the BW formula with the parameters from~\cite{pdg04},
and obtain $N_{sig}(D_{sJ}(2573))=7.7\pm 5.7$,
which corresponds to
the $90\%$ C.L. upper limit of ${\cal B} (B^{+}\to \bar{D}^{0}D_{sJ}(2573))\times 
{\cal B}(D_{sJ}(2573)\to D^{0}K^{+}) < 0.7 \times 10^{-4}$.

In summary, from a study of the Dalitz plot we find that the decay  
$B^{+}\to \bar{D}^{0}D^{0}K^{+}$ proceeds dominantly via
quasi-two-body channels: 
$B^{+} \to \bar{D}^{0} D_{sJ}^{+}(2700)$ and
$B^{+} \to \psi(3770) K^{+}$, 
where $D_{sJ}^{+}(2700)$ is a previously unobserved
resonance in the $D^{0}K^{+}$ system with a mass 
$M=2715 \pm 11 ^{+11}_{-14}~\rm{MeV}/c^{2}$, 
width  $\Gamma = 115 \pm 20 ^{+36}_{-32} ~\rm{MeV}/c^{2}$
and $J^P=1^-$.
The observed rate for $\psi(3770)$ production
confirms our previous observation~\cite{chistov}.

Based on its observed decay channel,
we interpret the  $D_{sJ}^{+}(2700)$
resonance as a $c\bar{s}$ meson.
Potential model calculations~\cite{godfrey-isgur} predict a $c\bar{s}$
radially excited  $2^{3}S_{1}$ state with mass $M=2720~\rm{MeV}/c^{2}$.
From chiral symmetry considerations~\cite{maciek} a $1^{+}$-$1^{-}$ 
doublet of states  has been predicted.  If the $1^+$ state
is identified as the $D_{sJ}(2536)$, the mass predicted for the $1^-$ state
is $M= 2721\pm 10~\rm{MeV}/c^{2}$. 
Additional measurements of the meson properties are needed to distinguish between
these two interpretations.

We thank the KEKB group for the excellent operation of the
accelerator, the KEK cryogenics group for the efficient
operation of the solenoid, and the KEK computer group and
the National Institute of Informatics for valuable computing
and Super-SINET network support. We acknowledge support from
the Ministry of Education, Culture, Sports, Science, and
Technology of Japan and the Japan Society for the Promotion
of Science; the Australian Research Council and the
Australian Department of Education, Science and Training;
the National Science Foundation of China and the Knowledge
Innovation Program of the Chinese Academy of Sciencies under
contract No.~10575109 and IHEP-U-503; the Department of
Science and Technology of India; 
the BK21 program of the Ministry of Education of Korea, 
the CHEP SRC program and Basic Research program 
(grant No.~R01-2005-000-10089-0) of the Korea Science and
Engineering Foundation, and the Pure Basic Research Group 
program of the Korea Research Foundation; 
the Polish State Committee for Scientific Research; 
the Ministry of Science and Technology of the Russian
Federation; the Slovenian Research Agency;  the Swiss
National Science Foundation; the National Science Council
and the Ministry of Education of Taiwan; and the U.S.\
Department of Energy.


%


\begin{thebibliography}{99}
\bibitem{cc}
Throughout this paper, 
the inclusion of the charge conjugate mode decay is implied.
\end{thebibliography}
\end{document}